\documentclass[a4paper,fleqn,usenatbib,useAMS]{mnras}
\usepackage{graphics}
\usepackage{graphicx}
\usepackage[pagewise]{lineno}
\usepackage{amsmath}    
\usepackage{amssymb}    
\usepackage{multicol}        
\usepackage{bm}     
\usepackage{pdflscape}  
\usepackage{xcolor}
\usepackage{ae,aecompl}
\usepackage{subfigure}
\usepackage{scalerel}
\usepackage[english]{babel}
\usepackage{times}
\usepackage{physics}
\usepackage{etoolbox}
\usepackage{multirow, booktabs}
\usepackage{siunitx}
\usepackage{paralist}
\usepackage{makecell}
\usepackage{tabularx}
\usepackage{blindtext}
\usepackage{enumitem}
\usepackage[english]{babel}
\newlist{subquestion}{enumerate}{1}
\setlist[subquestion,1]{label=(\alph*)}
\usepackage{paralist}
\newcommand{\wfirst}{\textit{Roman}}

\title[Numerically studying extreme finite-source microlensing degeneracy]{Numerically studying the degeneracy problem in extreme finite-source microlensing events}

\author[Sajadian]{Sedighe Sajadian$~^{1}$\thanks{E-mail: s.sajadian@iut.ac.ir}\\
$^{1}$Department~of~Physics,~Isfahan~University~of~Technology,~Isfahan~84156-83111,~Iran}

\begin{document}
\label{firstpage}
\pagerange{\pageref{firstpage}--\pageref{lastpage}}
\maketitle
\begin{abstract}	
Most transit microlensing events due to very low-mass lens objects suffer from extreme finite-source effects. While modeling their light curves, there is a known continuous degeneracy between their relevant lensing parameters, i.e., the source angular radius normalized to the angular Einstein radius $\rho_{\star}$, the Einstein crossing time $t_{\rm E}$, the lens impact parameter $u_{0}$, the blending parameter, and the stellar apparent magnitude. In this work, I numerically study the origin of this degeneracy. I find that these light curves have 5 observational parameters (i.e., the baseline magnitude, the maximum deviation in the magnification factor, the Full Width at Half Maximum $\rm{FWHM}=2 t_{\rm{HM}}$,  the deviation from top-hat model, the time of the maximum time-derivative of microlensing light curves $T_{\rm{max}}=t_{\rm E}\sqrt{\rho_{\star}^{2}-u_{0}^{2}}$). For extreme finite-source microlensing events due to uniform source stars we get $t_{\rm{HM}}\simeq T_{\rm{max}}$, and the deviation from the top-hat model tends to zero which both cause the known continuous degeneracy. When either $\rho_{\star}\lesssim10$ or the limb-darkening effect is considerable $t_{\rm{HM}}$, and $T_{\rm{max}}$ are two independent observational parameters. I use a numerical approach, i.e., Random Forests containing $100$-$120$ Decision Trees, to study how these observational parameters are efficient in yielding the lensing parameters. These machine learning models find the mentioned 5 lensing parameters for finite-source microlensing events from uniform, and limb-darkened source stars with the average $R^{2}$-scores of $0.87$, and $0.84$, respectively. $R^{2}$-score for evaluating the lens impact parameter gets worse on adding limb darkening, and for extracting the limb-darkening coefficient itself this score falls as low as $0.67$.   
\end{abstract}

\section{Introduction}
An important issue in microlensing observations \citep{Einstein1936,Liebes1964,Chang1979,Paczynski86} and analyzing their photometry light curves is degeneracy while modeling and finding the lensing parameters. There are two reasons for microlensing degeneracies as explained in the following. 
\begin{itemize}
\item Sometimes several different sets of relevant lensing parameters (e.g., the lensing time scale $t_{\rm E}$, the lens impact parameter $u_{0}$, the source radius, ...) generate similar microlensing light curves either completely accidentally \citep[see, e.g., ][]{1998ApJGaudi,2009ApJHAn}, or because of symmetry in the lensing formalism (either the lens equation or the magnification factor) with respect to special changes in relevant parameters \citep[][]{1999DominikAA,1994ApJGould,1997GaudiGould}.
	
\item On the other hand, usually the measured lensing parameters are not sufficient to evaluate the relevant physical parameters of the lens and source stars (e.g., the lens mass, the source radius, the lens distance, etc). To model light curves and infer these physical parameters, people usually use the Bayesian analysis to specify the physical parameters according to lensing parameters of microlensing light curves \citep[see, e.g., ][]{2010AACassan,2022AAHan}.
\end{itemize}

From another point of view, microlensing degeneracies can be either discrete or continuous \citep[e.g., see ][]{2022Johnson,2022NatAsZhang}. As a simple example of discrete degeneracies, all microlensing light curves do not change under the conversion $u_{0}$ to -$u_{0}$ (where, $u_{0}$ is the lens impact parameter). This kind of degeneracies is usually owing to symmetry in either amplification relations or lens equations under some conversions \citep[see, e.g., ][]{1998GriestApJ,1999DominikAA}.

Continuous degeneracies occurs when very small variations in several relevant lensing parameters do not change light curves' shape. In these cases, the number of observational parameters from microlensing light curves is less than the number of involved lensing parameters. These lensing parameters in turn are functions of physical parameters of the lens and source stars \citep[see, e.g.,][]{1997Paczynski}. For instance, caustic-crossing features in binary microlensing events can be modeled by choosing different values of the source angular radius normalized to the angular Einstein radius $\rho_{\star}$, the mass ratio $q$, and the relative lens-source angular velocity $\mu_{\rm{rel}}$ in small ranges \citep{1997GaudiGould,2012Gaudireview}. In these cases, sometimes by doing other measurements (e.g., polarimetric or astrometric observations) people can break these degeneracies \citep[see, e.g.,][]{2000GouldApJ,2002ApJAn,2017UnivLee,2014sajadian,2015MNRASsajadian,2022Sahu}. 

Another example of continuous degeneracies occurs in extreme finite-source and short-duration microlensing events which are mostly due to free-floating planets or low-mass lens objects. Several examples of such finite-source events due to low-mass lens objects have been reported recently \citep[e.g., ][]{2018MrozAJ,2019MrozAA, 2020MrozAJ,2020AJHan,2021AJKim,2021AJRyu}. Resolving this degeneracy is important because the \textit{The Nancy Grace Roman Space Telescope} (\wfirst)~telescope has been planned to detect a considerable number of short-duration microlensing events during its Galactic Bulge Time Domain Survey \citep{2019ApJSPenny,2020AJJohnson,2019MNRASbagheri}. 
\begin{figure*}
\includegraphics[width=0.33\textwidth]{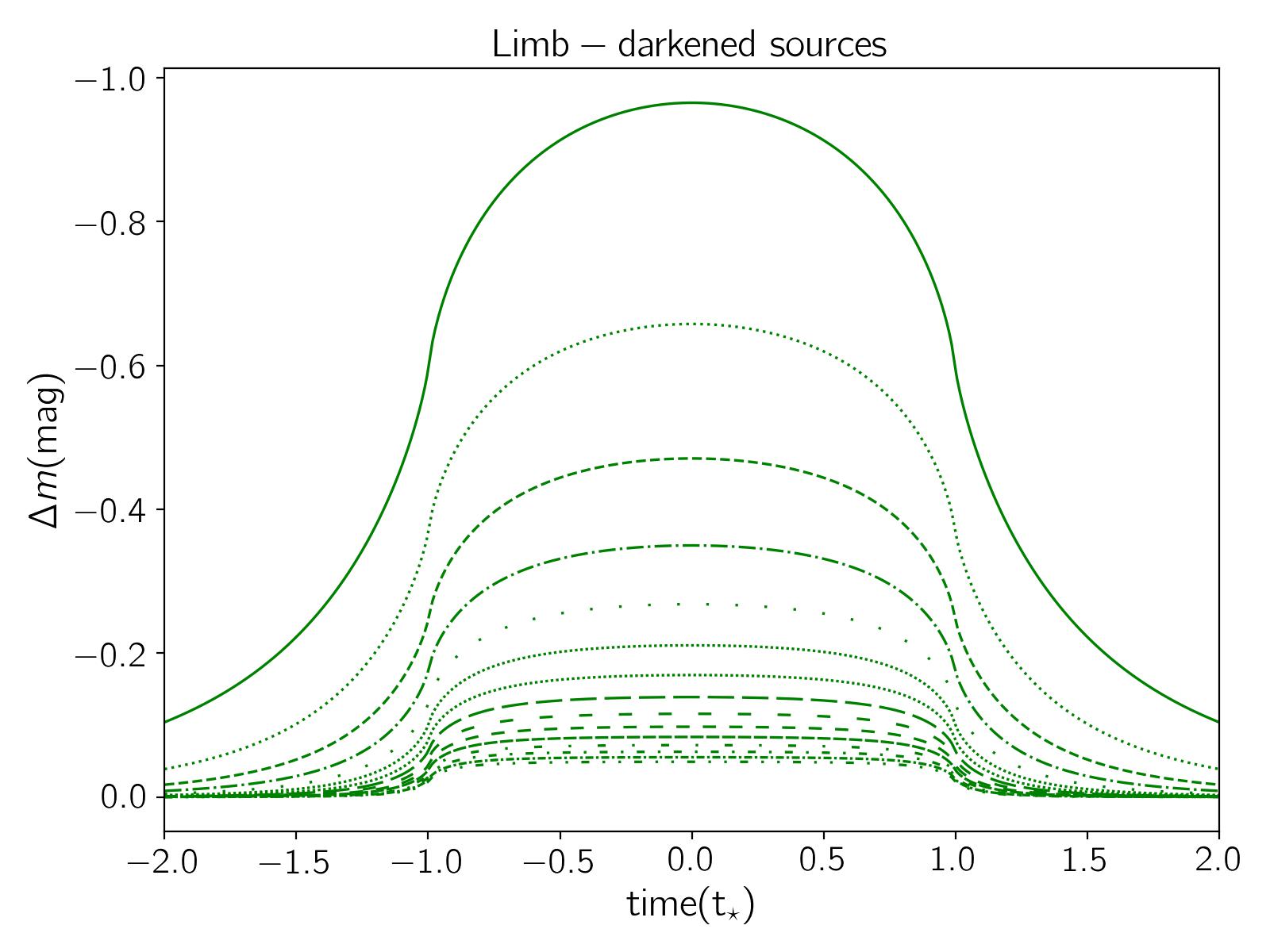}
\includegraphics[width=0.33\textwidth]{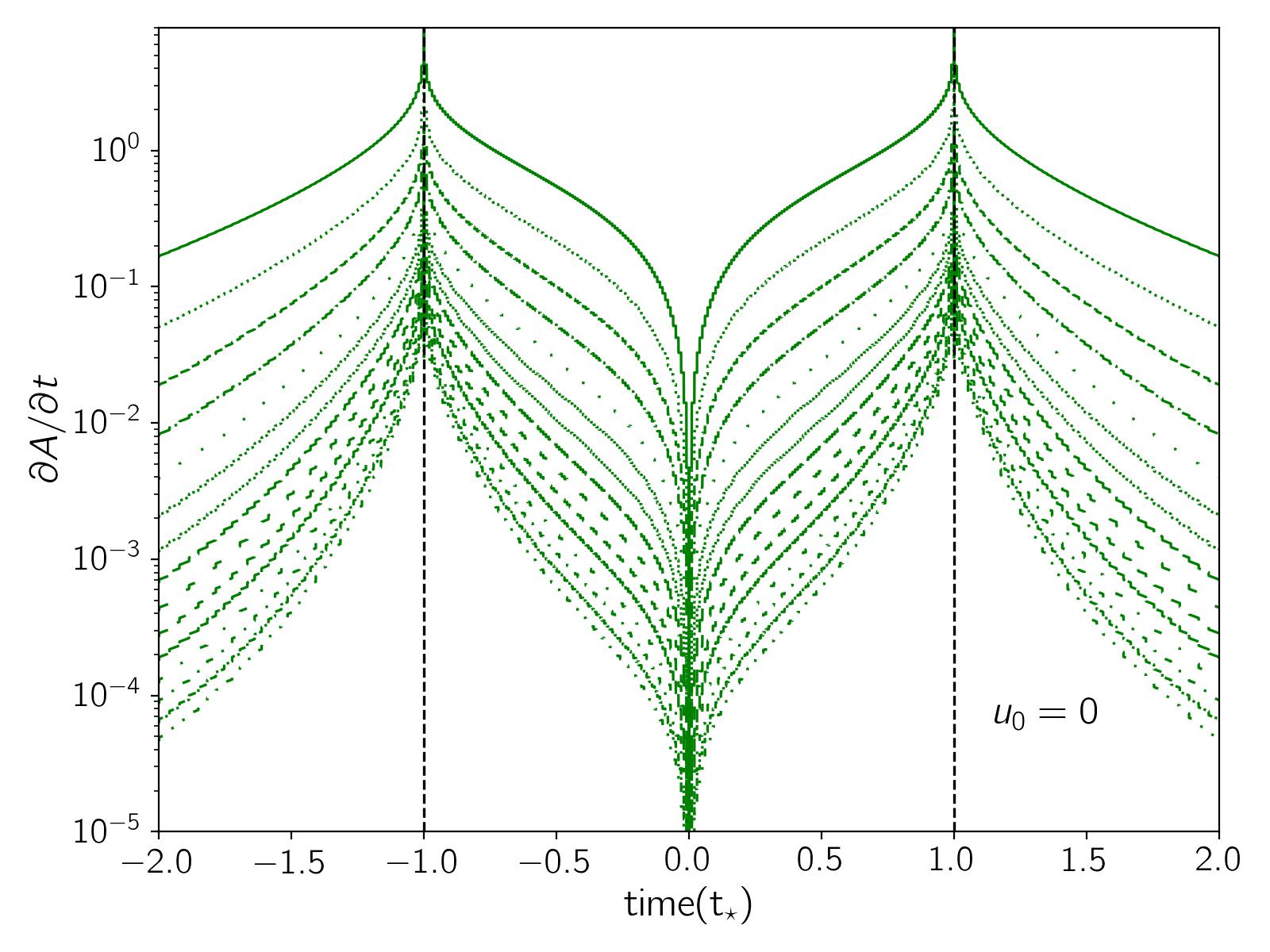}
\includegraphics[width=0.33\textwidth]{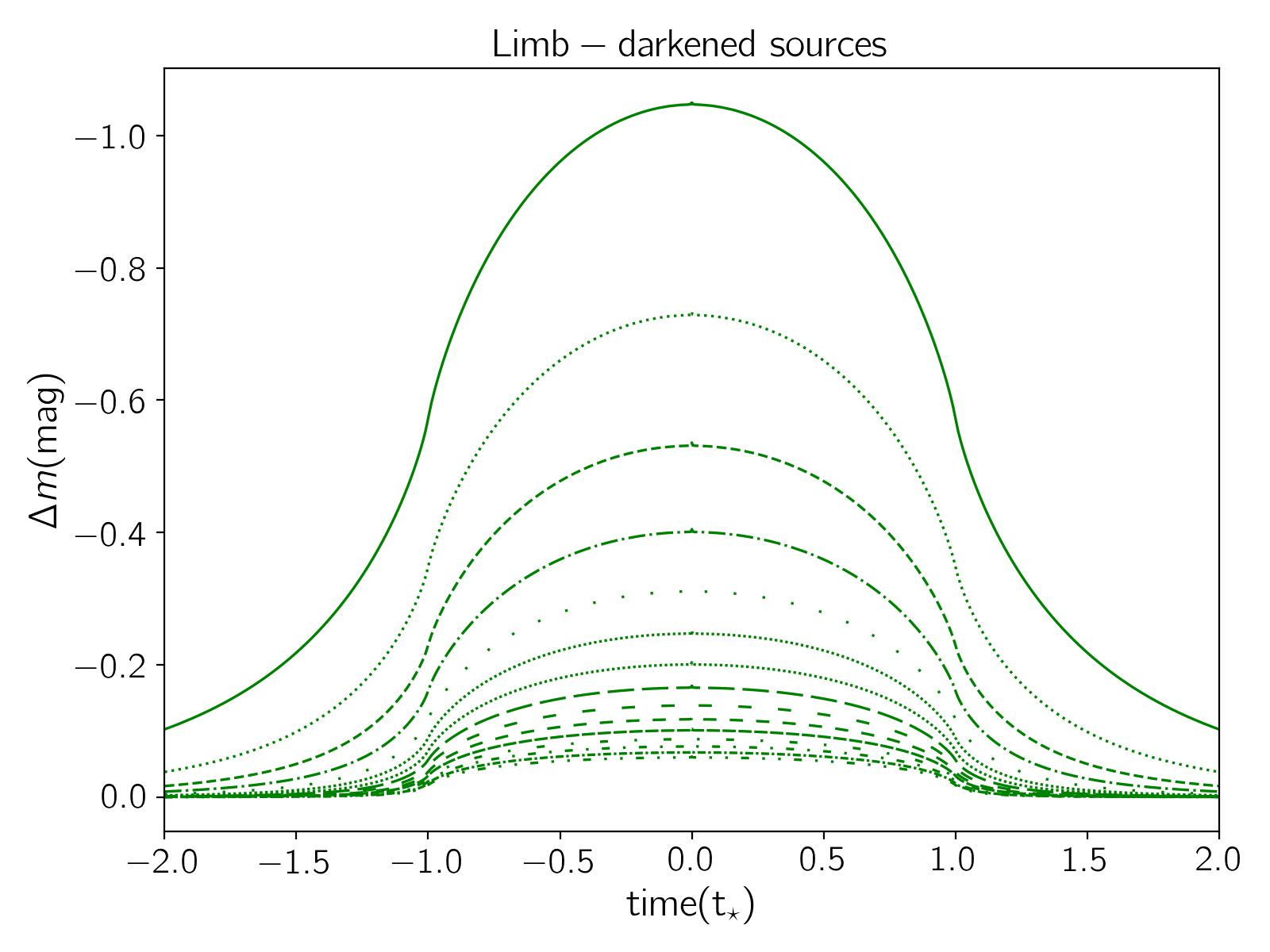}
\caption{Examples of finite-source microlensing events from uniform (first panel) and limb-darkened (last panel) source stars. The lensing parameters of these light curves are $\rho_{\star} = 0.9,~1.3,~1.7,~2.1,...,~6.5$, $u_{0} =0.1\rho_{\star}$, $t_{\star}=1$ day (where, $t_{\star}=t_{\rm E}~\rho_{\star}$), and $f_{\rm b}=1$, and $\Gamma \in 0.6$. In the middle panel, I represent the time-derivative of several finite-source microlensing light curves (Equation \ref{derieq}). For microlensing events in this panel, I set $u_{0}=0$ to have continuous curves. The black dashed vertical lines are plotted at $t=t_{\star}$.}\label{light} 
\end{figure*}

In such events, the relevant lensing parameters in the magnification factor are (i) $\rho_{\star}$, (ii) $u_{0}$, (iii) $t_{\rm E}$ (the Einstein crossing time), (iv) $t_{0}$ (the time of the closest approach), (v) $f_{\rm b}$ (the blending parameter which is the fraction of the source flux to the baseline flux), and (vi) $m_{\star}$ (the apparent magnitude of the source star without any lensing effect). $t_{0}$ can be directly extracted from microlensing light curves. Hence, by excluding $t_{0}$, there are five relevant lensing parameters. 

\noindent A microlensing light curve has several observational parameters (i.e., the parameters that are extracted from light curves's shape) which are functions of these lensing parameters. These parameters are (i) Full Width at Half Maximum (FWHM), (ii) the maximum deviation in the magnification factor $\Delta A$, (iii) the baseline apparent magnitude $m_{\rm{base}}$, (iv) the deviation from a top-hat model $\Delta_{\rm{TH}}$, as proposed by \citet{2022Johnson}.
 
\noindent Comparing the numbers of lensing parameters and observational parameters, the lensing parameters can not be determined uniquely. In this paper, I introduce another observational parameter for finite-source microlensing events which is the time of the maximum time-derivative of microlensing light curves, $T_{\rm{max}}$, and study if this new parameter is useful for resolving microlensing degeneracy. At this time ($T_{\rm{max}}$) the
 variation of stellar color maximizes as well \citep[see, Fig. (5) of ][]{2022AAsajadian}. Hence, microlensing observations in two filters simultaneously might specify this observational parameter. 

This time is a function of lensing parameters, as $T_{\rm{max}}=t_{\rm E}\sqrt{\rho_{\star}^{2}-u_{0}^{2}}$. For extreme finite-source events (i.e., $\rho_{\star}\gtrsim 10$) from uniform source stars the time of half maximum ($t_{\rm{HM}}$) is equal to $T_{\rm{max}}$ which results the known continuous degeneracy. For either limb-darkened source stars, or when $\rho_{\star}\lesssim10$, these times, $T_{\rm{max}}$ and $t_{\rm{HM}}$, are not similar and both of them help resolving this microlensing degeneracy.  

For a large sample of finite-source microlensing events, I numerically study how these observational parameters are efficient to determine the relevant lensing parameters by the aid of Supervised Machine Learning approaches. The Machine Learning (ML) models provide abilities for computers to learn from data, find constructions of data, create a model, and finally forecast outputs for unseen inputs. When we do not have analytical relations between input and output parameters, ML methods can automatically find correlations between several inputs and outputs, classifying inputs into discrete classes, finding a function between different inputs, etc. \citep[e.g., ][]{james2013introduction,chattopadhyay2014statistical}. Hence, the ML approaches could find relations between observational parameters (inputs) and lensing parameters (outputs) for finite-source microlensing events.

The outline of the paper is as follows. In Section \ref{finite}, I first review the known formalism for finite-source microlensing events. Then, in Subsections \ref{feature}, and \ref{limbd}, I numerically investigate the relations between observational parameters and the lensing parameters for these light curves due to uniform and limb-darkened source stars, respectively. The correlation coefficients between the observational and lensing parameters are reported in Subsection \ref{correlation}. In Section, \ref{tree}, I numerically examine the efficiency of these observational parameters to determine the relevant lensing parameters using several Random Forests contain 100-120 Decision Trees. In Section \ref{conclu}, I summarize the results and conclude.

\section{Finite-source microlensing events}\label{finite}
Low-mass lens objects, such as free-floating or wide-orbital planets, make short-duration microlensing events, that can potentially be discovered through dense observations toward crowded fields in the Galactic bulge \citep{2011NaturSumi,2017NaturMroz,2019ApJSPenny}. For these events, the typical values of the angular Einstein radius $\theta_{\rm E}$ and the normalized angular source radius $\rho_{\star}$ are (respectively):    
\begin{eqnarray}
\theta_{\rm E}(\mu \rm{as})&=& 1.0~\sqrt{\frac{M_{\rm l}}{M_{\oplus}}~\frac{\pi_{\rm{rel}}\rm{(mas)} }{0.04} },\nonumber\\
\rho_{\star} &=& 0.58~\frac{\theta_{\star}(\mu \rm{as})}{\theta_{\odot}}~\frac{1.0}{ \theta_{\rm E}(\mu \rm{as})},
\end{eqnarray}
\noindent where, $M_{\rm l}$ is the lens mass, and $\pi_{\rm{rel}}=\rm{au}\big(1/D_{\rm l} -1/D_{\rm s}\big)$ is the relative parallax amplitude. By assuming $D_{\rm l}=6$ kpc, and $D_{\rm s}=8$ kpc, we will have $\pi_{\rm{rel}}=0.04$ mas. $\theta_{\star}$ is the angular source radius, which for a Sun-like source star inside the Galactic bulge is $\theta_{\odot}=0.58$ $\mu$as. Accordingly, for these events the finite-source effect \citep{1994WittMOA} is considerable \citep[see, e.g., ][]{2021MNRASSAjadian}. Comparing to most-common microlensing events due to red dwarf objects with $M_{\rm l}\simeq 0.3~M_{\odot}$ \citep[e.g., ][]{2006MNRASDominik}, $\theta_{\rm E}$ for these short-duration events are smaller by a factor $300$, which yields a larger $\rho_{\star}$ ($\sim 300$ times). 

The magnification factor in finite-source microlensing events is given by \citep{1994WittMOA}:
\begin{eqnarray}
A(u,~\rho_{\star})&=&\frac{1}{\pi}\Big[-\frac{u-\rho_{\star}}{\rho_{\star}^{2}}\frac{8+u^{2}-\rho_{\star}^{2}}{\sqrt{4+(u-\rho_{\star})^{2}}}~F\big(\frac{\pi}{2}, k\big)\nonumber\\ &+& \frac{u+\rho_{\star}}{\rho_{\star}^{2}}\sqrt{4+(u-\rho_{\star})^{2}}~E\big(\frac{\pi}{2}, k\big)\nonumber\\&+&\frac{4(u-\rho_{\star})^{2}}{\rho_{\star}^{2}(u+\rho_{\star})}\frac{1+\rho_{\star}^{2}}{\sqrt{4+(u-\rho_{\
			\star})^{2}}}~\Pi\big(\frac{\pi}{2}, n,k\big) \Big], 
\end{eqnarray}\label{magni}
where, $u$ is the lens-source distance projected on the lens plane and normalized to the Einstein radius (i.e., radius of the images ring at the complete alignment), $n=4~u~\rho_{\star}\big/(u+\rho_{\star})^{2}$, and $k=\sqrt{\frac{4~n}{4~+~(u-\rho_{\star})^{2}}}$. The functions $F$,  $E$, and $\Pi$ are the first, second and third types of the elliptical integral, respectively.

Several examples of finite-source microlensing events from uniform source stars due to low-mass lens objects (with different values of $\rho_{\star}$) are shown in the first panel of Figure \ref{light}. For generating these microlensing light curves, we choose the lensing parameters as $\rho_{\star} = 0.9,~1.3,~1.7,~2.1,...,~6.5$, $u_{0} =0.1~\rho_{\star}$, $t_{\star}=1$ day, and $f_{\rm b}=1$. The $y$-axis shows the variation in stellar apparent magnitude which is $\Delta m(\rm{mag})= -2.5 \log_{10}\big[f_{\rm b} A(u,~\rho_{\star}) +1-f_{\rm b}\big]$. Here, $A(u,~\rho_{\star})$ is the magnification factor as given in Equation \ref{magni}. For calculating finite-source magnification factor, we use the RT-model \footnote{\url{http://www.fisica.unisa.it/GravitationAstrophysics/VBBinaryLensing.htm}} which was well developed by V.~Bozza \citep{2010MNRASBozza,2018MNRASBozza}. Accordingly, the finite-source effect makes light curves be flattened.

\subsection{Observational parameters in finite-source microlensing}\label{feature}
A microlensing light curve with finite-source effect from a uniform source star (without limb-darkening effect) is a function of five lensing parameters, including $t_{\rm E}$, $u_{0}$, $\rho_{\star}$, $f_{\rm b}$, and $m_{\star}$. We exclude the time of the closest approach ($t_{0}$) because it will be measured from observations directly. We add the limb-darkening effect in Subsection \ref{limbd}. If the number of observational parameters (extracted from microlensing light curves) is less than five, there is a continuous degeneracy. There are several observational parameters for finite-source microlensing light curves, which are listed in the following.   
\begin{figure*}
\subfigure[]{\includegraphics[width=0.49\textwidth]{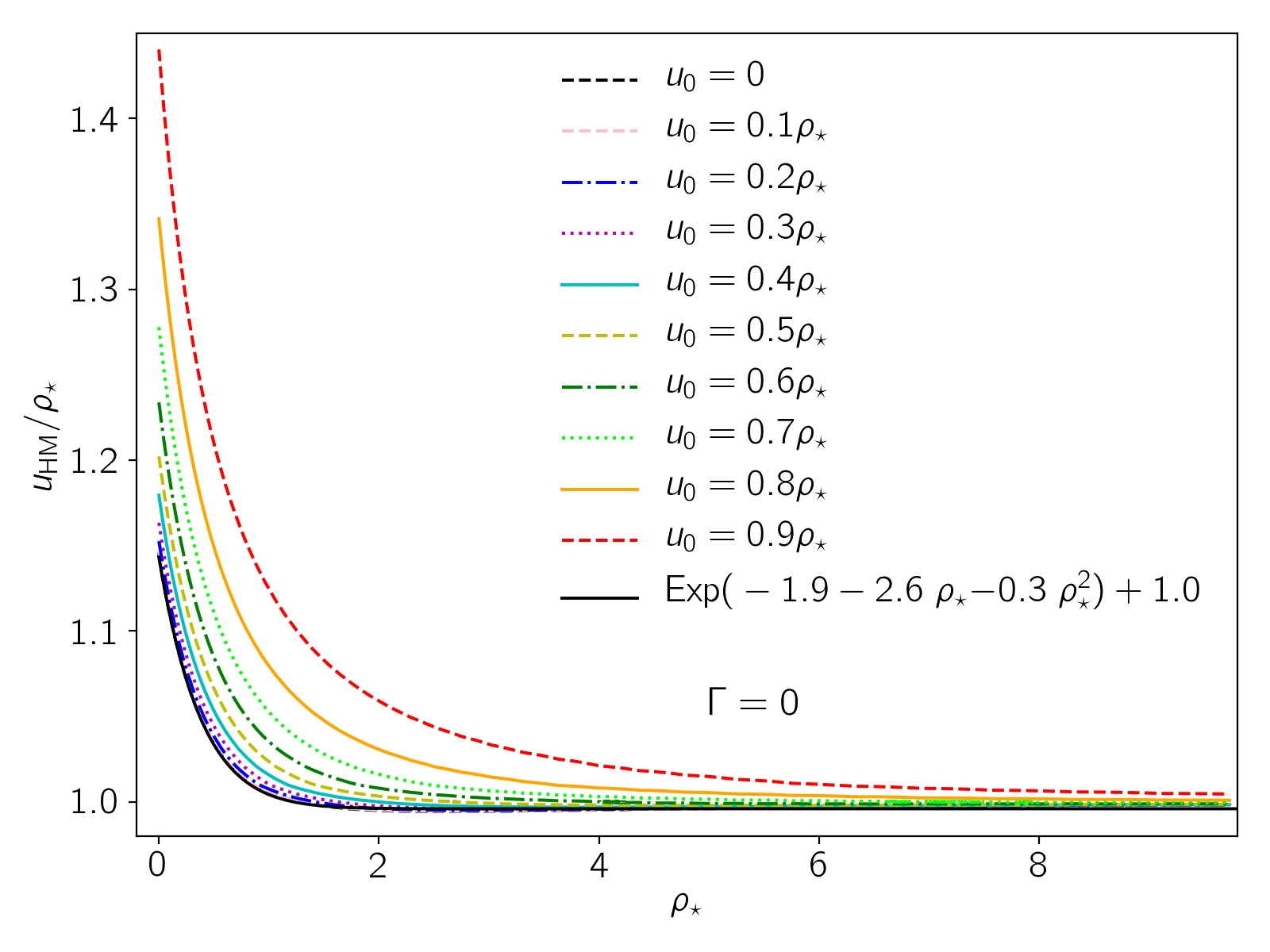}\label{uhm1}}
\subfigure[]{\includegraphics[width=0.49\textwidth]{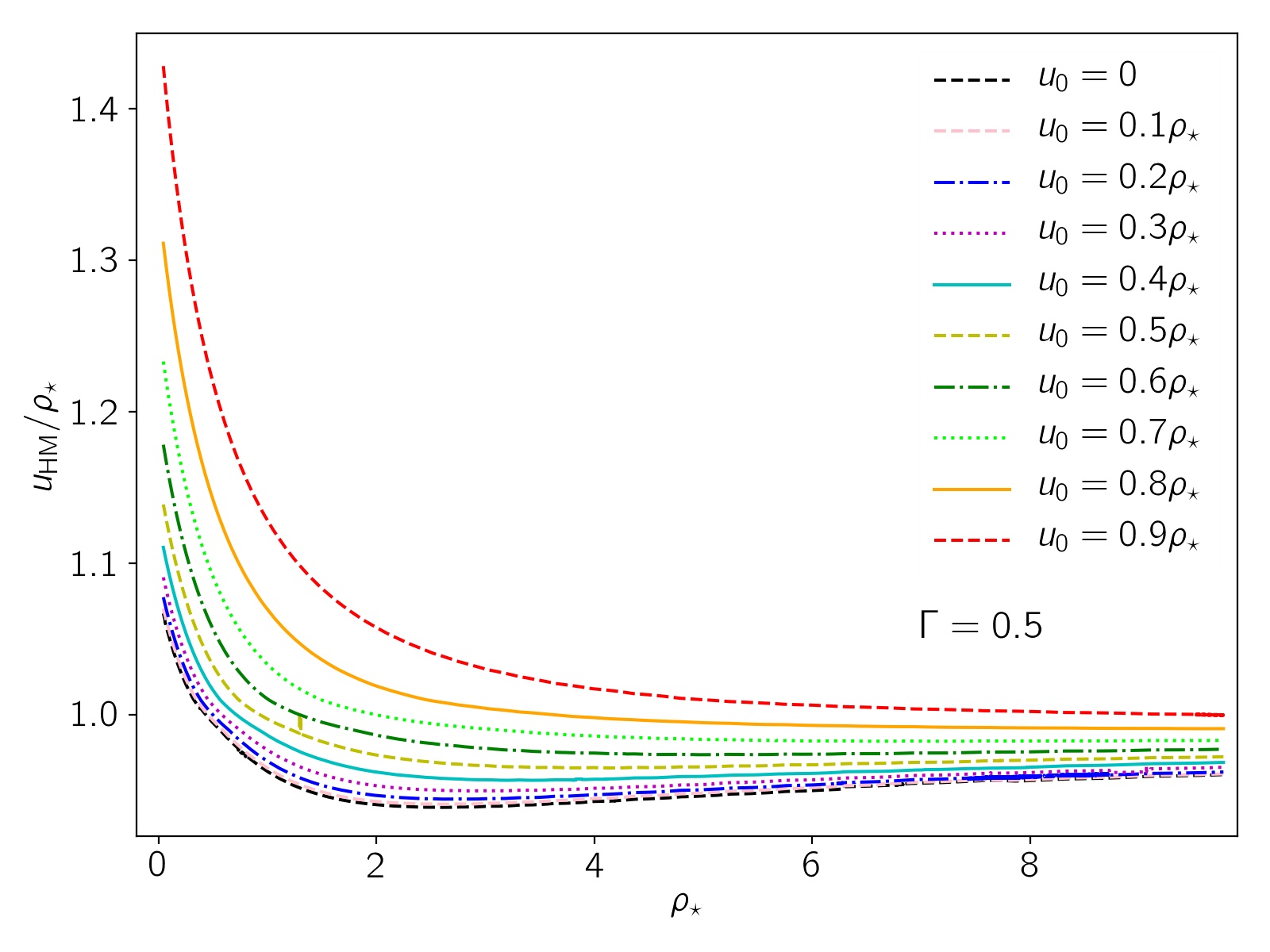}\label{uhm2}}
\subfigure[]{\includegraphics[width=0.49\textwidth]{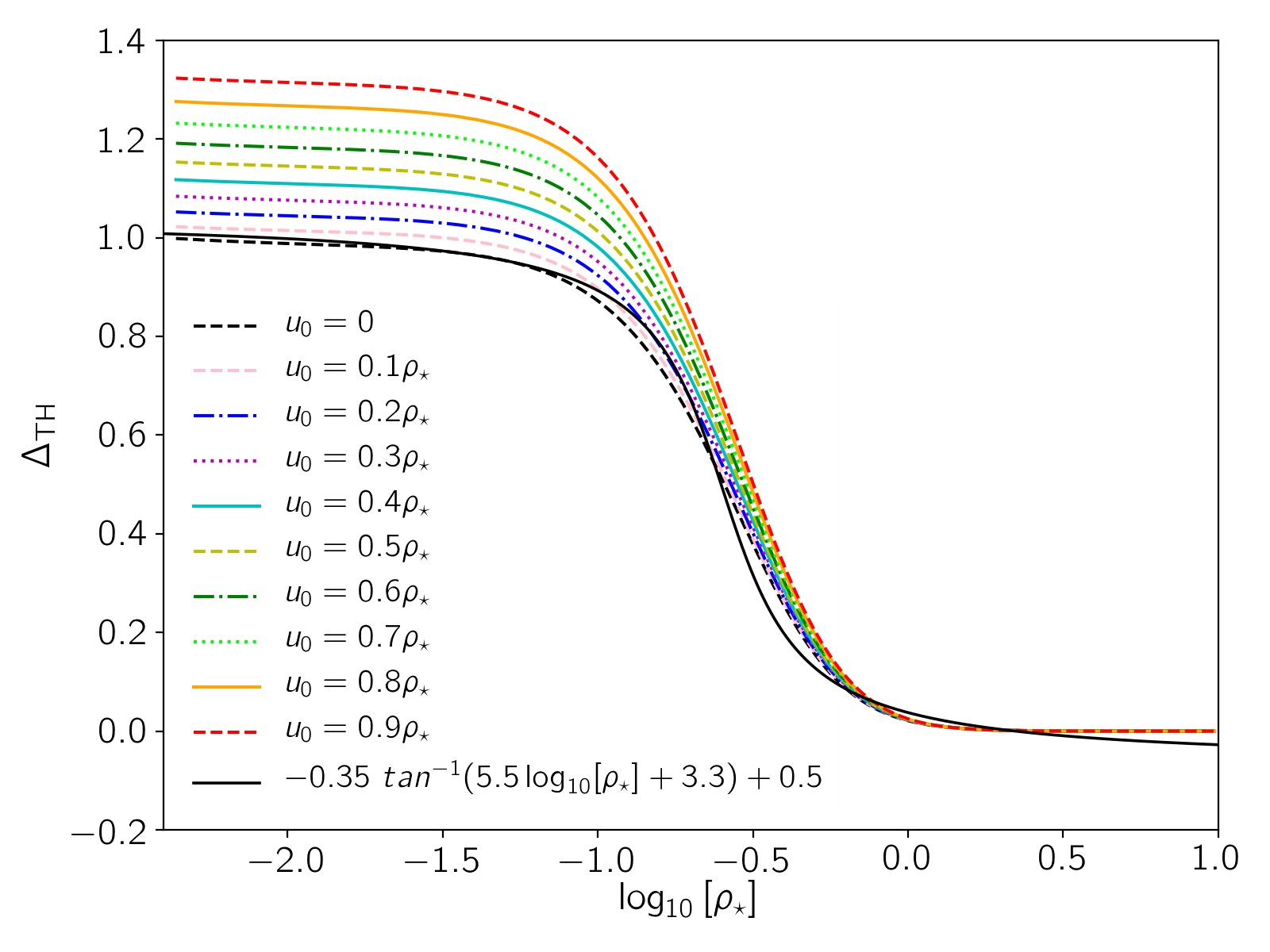}\label{devia}}
\subfigure[]{\includegraphics[width=0.49\textwidth]{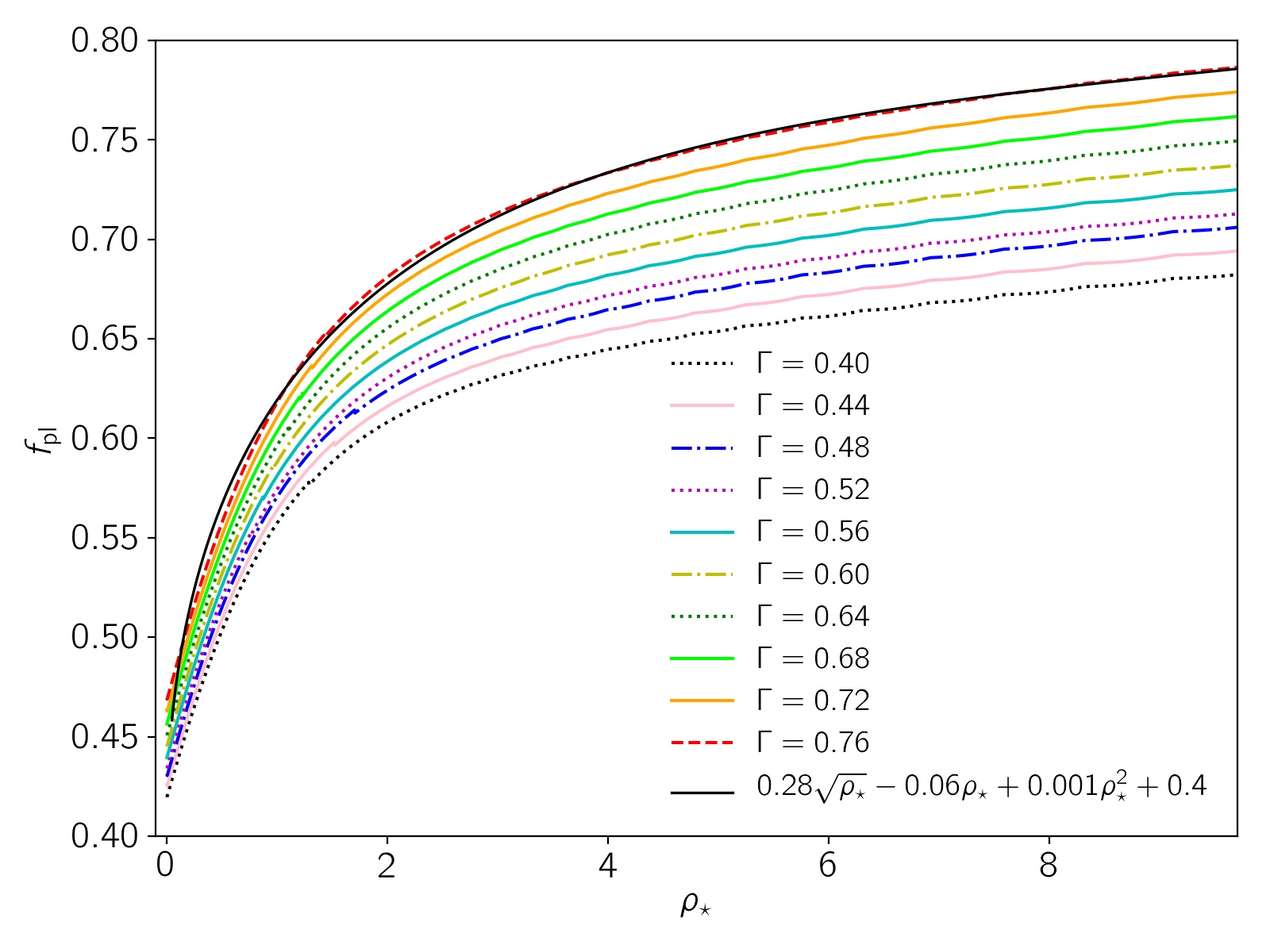}\label{fplfig}}	
\caption{Two top panels show the lens-source distance normalized to the Einstein radius at the half maximum, $u_{\rm{HM}}$, divided by $\rho_{\star}$ versus the normalized source radius for uniform ($\Gamma=0$) and limb-darkened ($\Gamma=0.5$) source stars. Two bottom panels represent $\Delta_{\rm{TH}}$ and $f_{\rm{pl}}$ for different values of the lens impact parameters versus $\rho_{\star}$.}
\label{uhm}
\end{figure*}

\begin{enumerate}
\item $\mathbf{m_{\rm{base}}}:$~The baseline apparent magnitude which is $m_{\rm{base}}=m_{\star}+2.5 \log_{10}\big[f_{\rm b}\big]$. It is proportional to the source apparent magnitude.\\

\item $\boldsymbol{\Delta A}:$~The maximum enhancement in the magnification factor with respect to the baseline which is $\Delta A= A(u_{0},~\rho_{\star})-1$. We note that for extreme finite-source microlensing events, $A(u_{0},~\rho_{\star})\sim 1+ 2~\rho_{\star}^{-2}$, and it depends strongly on $\rho_{\star}$ \citep{1997ApJGould}.\\  

\item \textbf{FWHM:}~The Full Width at Half Maximum of microlensing light curves, which is 
\begin{eqnarray}
\rm{FWHM}=2~t_{\rm{HM}}=2~t_{\rm E}\sqrt{u_{\rm{HM}}^{2} - u_{0}^{2}},
\end{eqnarray}
where, $u_{\rm{HM}}$ is the lens-source angular distance normalized to the angular Einstein radius at the half maximum. Since the magnification factor for finite-source microlensing (Equation \ref{magni}) is a function of elliptical integrals, I numerically calculate $u_{\rm{HM}}$ for different values of $u_{0}$, and $\rho_{\star}$ as plotted in \ref{uhm1}. Generally, for a given value of $u_{0}$, $u_{\rm{HM}}$ is a decreasing and exponential function versus $\rho_{\star}$. For $u_{0}=0$, $u_{\rm{HM}}$ versus $\rho_{\star}$ is given by: $$u_{\rm{HM}}=\rho_{\star}+ 0.15~\rho_{\star}~\exp\big(-2.59 \rho_{\star}-0.29 \rho_{\star}^{2}\big).$$ For other values of the lens impact parameters, the coefficients of best-fitted exponential curves are given in Table \ref{app1} of Appendix \ref{append1}. Accordingly, the FWHM of extreme finite-source microlensing light curves, i.e., $\rho_{\star} \gtrsim 10$, regardless of the lens impact parameter tends to a constant value, i.e., $\rm{FWHM}\simeq 2~t_{\star}$. Here $t_{\star}=t_{\rm E}~\rho_{\star}$ is the time scale of crossing the source radius by the lens.\\
 
\item $\mathbf{T_{\rm{max}}}:$~The next observational parameter is the time of the maximum $u$-derivative of the magnification factor, i.e., $A'$. This maximum occurs at $u_{\rm{max}}=\rho_{\star}$. 
Hence, the maximum of $A'$ happens at the time (with respect to the time of the closest approach):
\begin{eqnarray}
 T_{\rm{max}}= t_{\rm E}\sqrt{\rho_{\star}^{2}-u_{0}^{2}}.
\end{eqnarray}
We note that the $u$-derivative is related to time-derivative of the magnification factor as:
\begin{eqnarray}
A'=\frac{\partial A}{\partial u}= \frac{\partial A}{\partial t}\frac{u~t_{\rm E}}{\sqrt{u^{2}-u_{0}^{2}}}.
\label{derieq}
\end{eqnarray}
Since, the second factor in the above equation is always positive, so the maxima of both $A'$ and $\partial A/\partial t$ occur at $T_{\rm{max}}$.

\noindent In the middle panel of Figure \ref{light}, we show time-derivative curves of different finite-source microlensing light curves versus time normalized to $t_{\star}$, which are calculated numerically. For all of these curves I set $u_{0}=0$ to have continuous curves versus time. As mentioned, the maxima of these curves happen when $u = \rho_{\star}$ (or $t=t_{\star}$ if $u_{0}=0$).\\
 
\item $\boldsymbol{\Delta_{\rm{TH}}}:~$The deviation from a top-hat model is the next observational parameter. To evaluate this deviation, we introduce a top-hat model corresponding to each light curve, which is a heaviside step function, as
\begin{eqnarray}
A_{\rm{TH}}=\begin{cases}
A(u_0,~\rho_{\star})&~ u\leq u_{\rm{max}}, \\  
1 &~  u>u_{\rm{max}}. \\
\end{cases}
\end{eqnarray} 
We define the deviation of each light curve from its corresponding top-hat model as: 
\begin{eqnarray}
\Delta_{\rm{TH}}=\frac{1}{N}\sum_{i=0}^{N}\Big(\frac{A(u_{i})-A_{\rm{TH}}(u_{i}) }{A_{\rm{TH}}(u_{i})}\Big)^{2}
\end{eqnarray} 
This factor is the average of squared relative deviations in the magnification factor with respect to its corresponding top-hat model over time. $N$ is the number of $u_{i}$s. For different values of $\rho_{\star}$ and $u_{0}$, I calculate $\Delta_{\rm{TH}}$ numerically as depicted in Figure \ref{devia}. Accordingly, this observational parameter strongly depends on $\rho_{\star}$. By increasing the lens impact parameter, it enhances. The best-fitted model to $\Delta_{\rm{TH}}$ versus $\rho_{\star}$ for $u_{0}=0$ (which is an arctangent function, as given by Equation \ref{uhmd}) is displayed by a solid and black curve in this panel. For other values of the lens impact parameter, the coefficients of the best-fitted models are given in Table \ref{app2} of Appendix \ref{append1}. I note that for $\rho_{\star}\gtrsim 1$ this observational parameter is not a function of the lens impact parameter and, is very small, because the light curves' shape tends to a top-hat one.
\end{enumerate}

Accordingly, for extreme finite-source microlensing events from uniform source stars the degeneracy persists, because for these events we have $T_{\rm{max}}\sim t_{\rm{HM}}$, and the deviation from its corresponding top-hat model tends to zero. Hence, for these events the number of independent observational parameters will be three (FWHM, $\Delta A$, and $m_{\rm{base}}$) whereas the number of lensing parameters is five. This point justifies the reported continuous degeneracy while modeling such events \citep[see, e.g., ][]{2020MrozAJ,2022Johnson}. For other events, these five observational parameters have different dependencies to lensing parameters, and they can potentially resolve the degeneracy. 

\begin{figure*}
\includegraphics[width=0.49\textwidth]{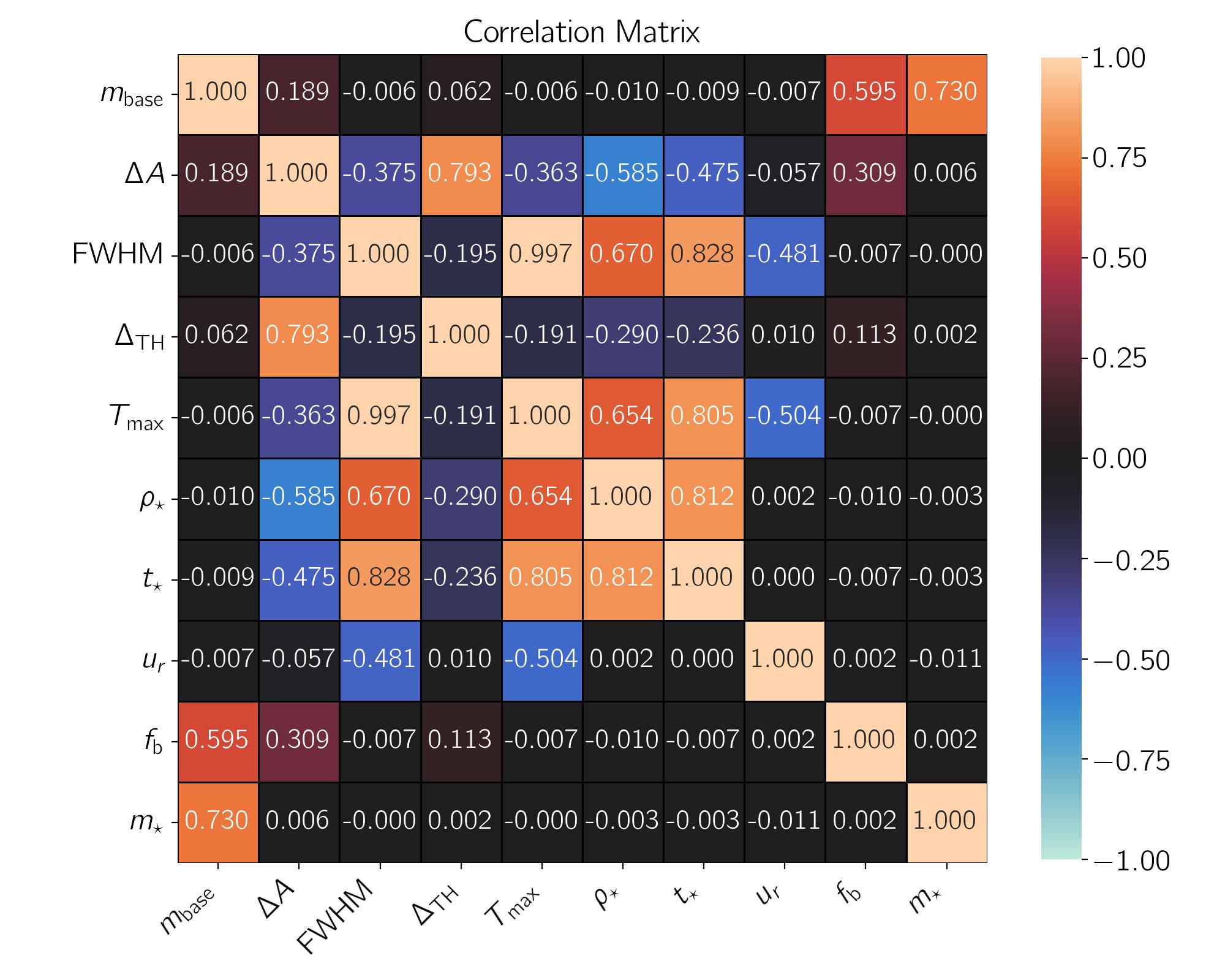}
\includegraphics[width=0.49\textwidth]{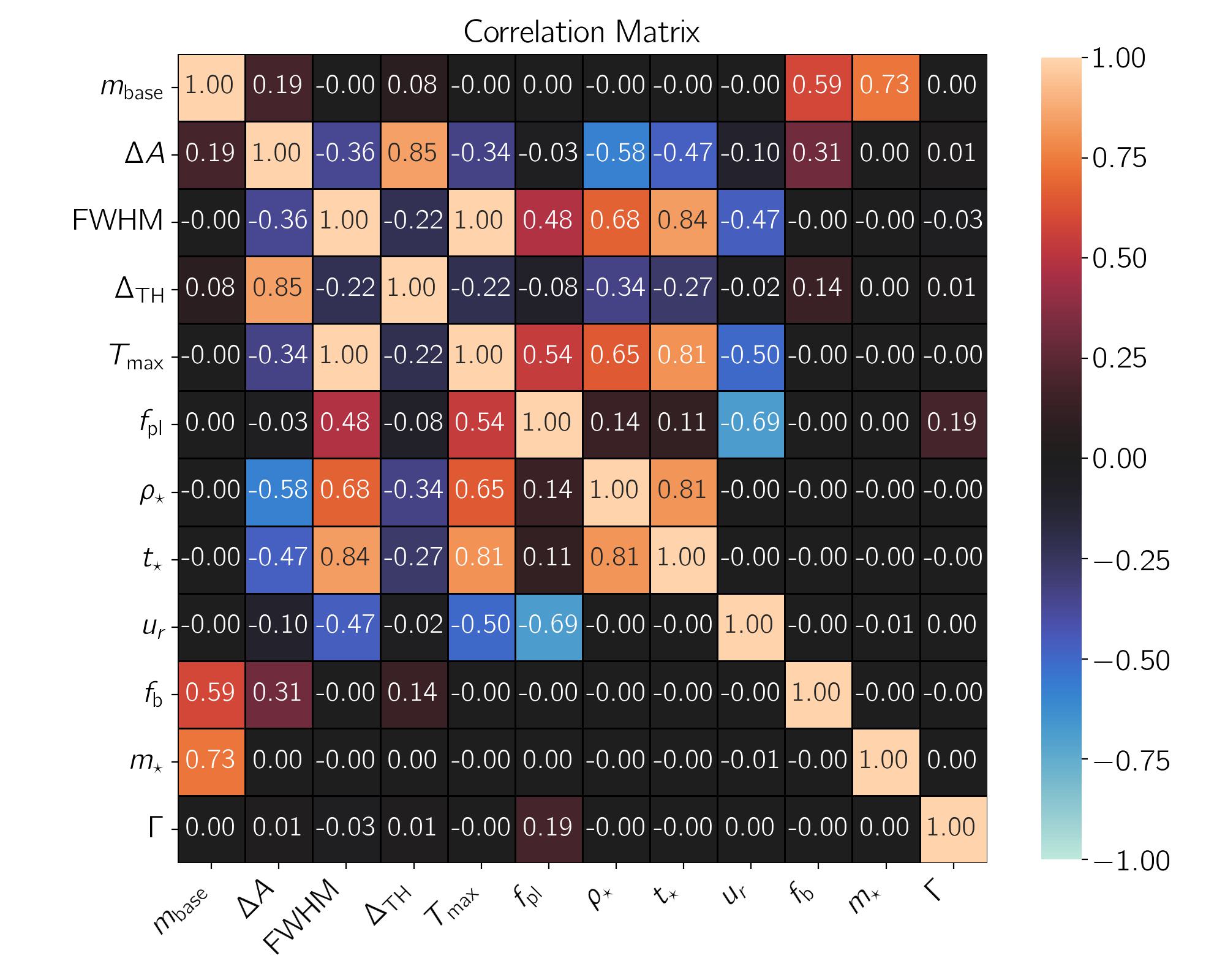}
\caption{The correlation matrices between observational parameters and lensing parameters as calculated by generating two large ensembles of finite-source microlensing events from uniform (left panel) and limb-darkened (right panel) source stars.}
\label{corri}
\end{figure*}
\subsection{Limb-darkened source stars}\label{limbd}
In reality the brightness profile over the stellar disk is not uniform, and it decreases from center to limb, the so-called limb-darkening effect. Generally, for a limb-darkened source star a simple and linear brightness profile is considered, i. e., $I(\mu)=I_{0}\big[1-\Gamma (1- \mu)\big]$, where $\mu=\sqrt{1-R^{2}\big/R_{\star}^{2}}$, $R$, and $R_{\star}$ are the radial distance from the source center over the stellar disk and the source radius, respectively. $\Gamma$ is the so-called limb-darkening coefficient. Examples of finite-source microlensing light curves due to limb-darkened source stars are represented in the last panel of Figure \ref{light}. By comparing the first and last panels, we note that the limb-darkening effect in finite-source and transit microlensing events diverts the light curves' shape from top-hat models. Therefore, in these events we include another observational parameter corresponding to the lensing parameter $\Gamma$, which is,
\begin{eqnarray}
f_{\rm{pl}}=\frac{A(u_{0},~\rho_{\star})-A(u_{\rm{max}},~\rho_{\star})}{A(u_{0},~\rho_{\star})}.
\label{fpl}
\end{eqnarray}
This observational parameter is the relative drop in the magnification factor from the time of the closest approach $t_{0}$ to the time of maximum time-derivative. In Figure \ref{fplfig} we show $f_{\rm{pl}}$ versus the source radius by considering different values for $\Gamma$, which is numerically calculated. Accordingly, this observational parameter depends on the limb-darkening coefficient even for large stellar radii. We also fit a power-law function to this observational parameter as $f_{\rm{pl}}=c_{1} \sqrt{\rho_{\star}} + c_{2} \rho_{\star} +c_{3} \rho_{\star}^{2} +c_{4}$. The coefficients of best-fitted models are given in Table \ref{app3} of Appendix \ref{append1}. 

Limb-darkening effect causes microlensing light curves to deviate from top-hat models, and it decreases the effective source radius. As a result, two observational parameters FWHM, and $T_{\rm{max}}$ behave independently. To probe this point, we plot $u_{\rm{HM}}$ for $\Gamma=0.5$ in Figure \ref{uhm2}. Comparing Figures \ref{uhm1}, and \ref{uhm2} the limb-darkening effect may resolve the lensing degeneracy for $\rho_{\star}\gtrsim 10$. 

In the next section, we study the correlations between these observational parameters and the lensing parameters through evaluating the correlation matrices.  

\subsection{Correlation matrix}\label{correlation}
In this section, we study the dependence of the observable parameters to the relevant lensing parameters through calculating their correlation matrix. First we make a big ensemble of finite-source and transit microlensing events. For generating these events, we need 5 lensing parameters which are chosen uniformly from ranges $u_{0} \in [0,~\rho_{\star}]$, $\rho_{\star} \in [1,~10]$, $t_{\rm E} \in [3,~7]$ days, $f_{\rm b} \in [0,~1]$, $m_{\star} \in [16,~19]$ mag, and $\Gamma \in [0.35,~0.7]$. We calculate the correlation matrices for observational and lensing parameters of simulated light curves as shown in Figure \ref{corri} without (left panel) and with (right panel) considering limb-darkening effects using \texttt{Numpy} Python module \footnote{\url{https://numpy.org/}}. We note that in these matrices $u_{\rm r}=u_{0} \big/ \rho_{\star}$. The correlation indices are numbers in the range of $[-1,~1]$. The zero correlation index between two given parameters means no dependency, $+1$, and $-1$ refer to linear relations with positive and negative slopes between two given entrances, respectively. Some key points from correlation matrices are listed in the following. 

\begin{itemize}
\item $m_{\rm{base}}$ is a function of $m_{\star}$, and $f_{\rm b}$, and has a linear relation (with a positive slope) with $m_{\star}$. 

\item $\Delta A$ has the highest correlation with $\rho_{\star}$. FWHM is proportional to $t_{\star}$, and has an inverse relation with $u_{\rm r}$.

\item The behavior of $T_{\rm{max}}$ is somewhat similar to that of FWHM. As explained in the previous section, for extreme finite-source microlensing events we have $T_{\rm{max}}\simeq t_{\rm{HM}}\simeq t_{\star}$.

\item The next observational parameter, i.e., $\Delta_{\rm{TH}}$, is significantly correlated by $\rho_{\star}$. 

\item The limb-darkening effect alters FWHM, $\Delta_{\rm{TH}}$, and $\Delta A$, because their correlation indices are not zero. The correlations of limb-darkening coefficient $\Gamma$ with all observational parameters are weak and $<0.2$.    
\end{itemize}  

Generally, the correlation indexes reveal that relation between observational parameters (e.g., FWHM, $T_{\rm{max}}$, ...)  are more complex than those are offered in Subsection \ref{feature}. Owing to lack of analytical relations between them, I test predicting the lensing parameters from the observational parameters numerically and using several Random Forests, as explained in the next section.
\begin{figure*}
\includegraphics[width=0.49\textwidth]{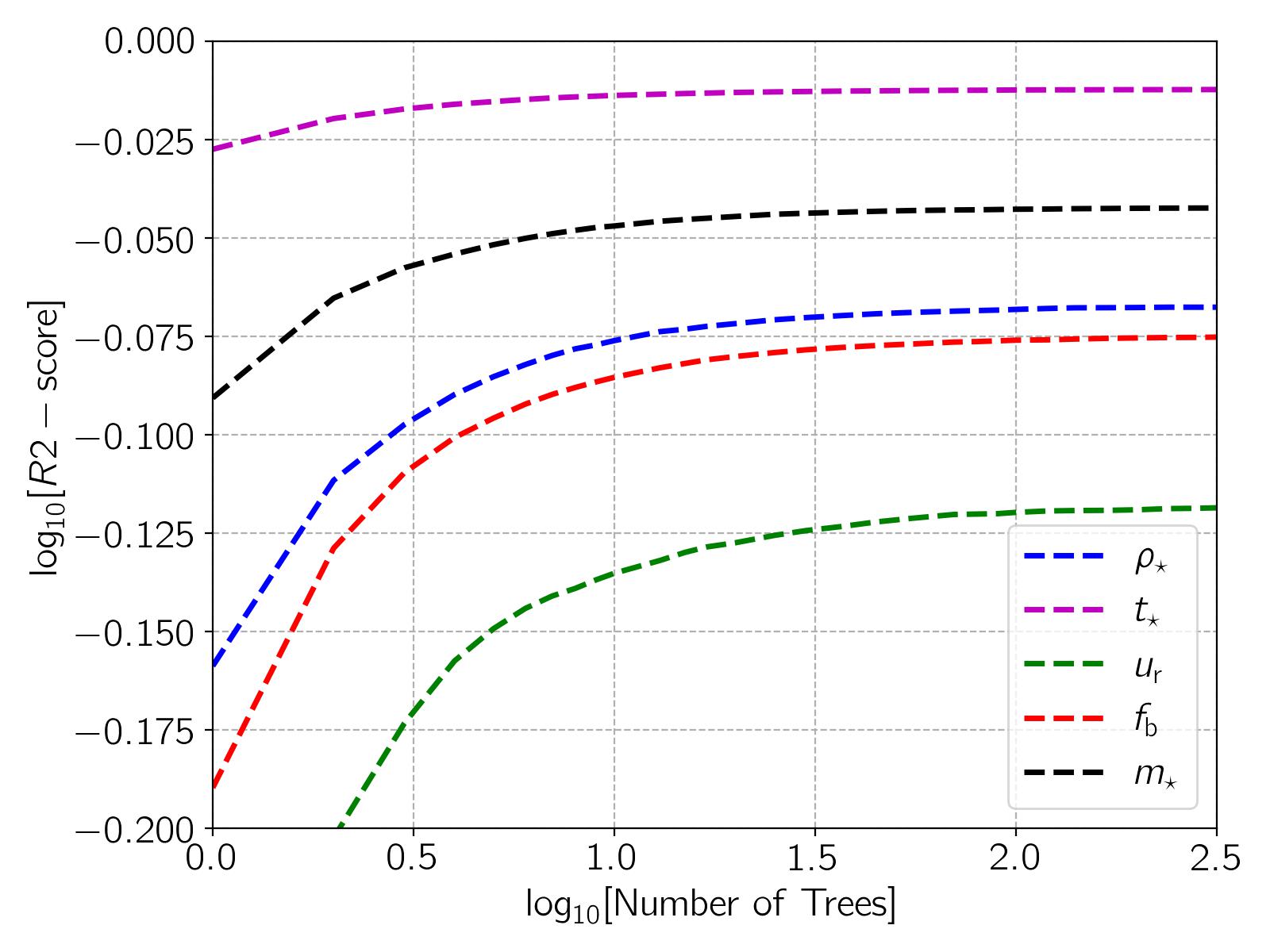}
\includegraphics[width=0.49\textwidth]{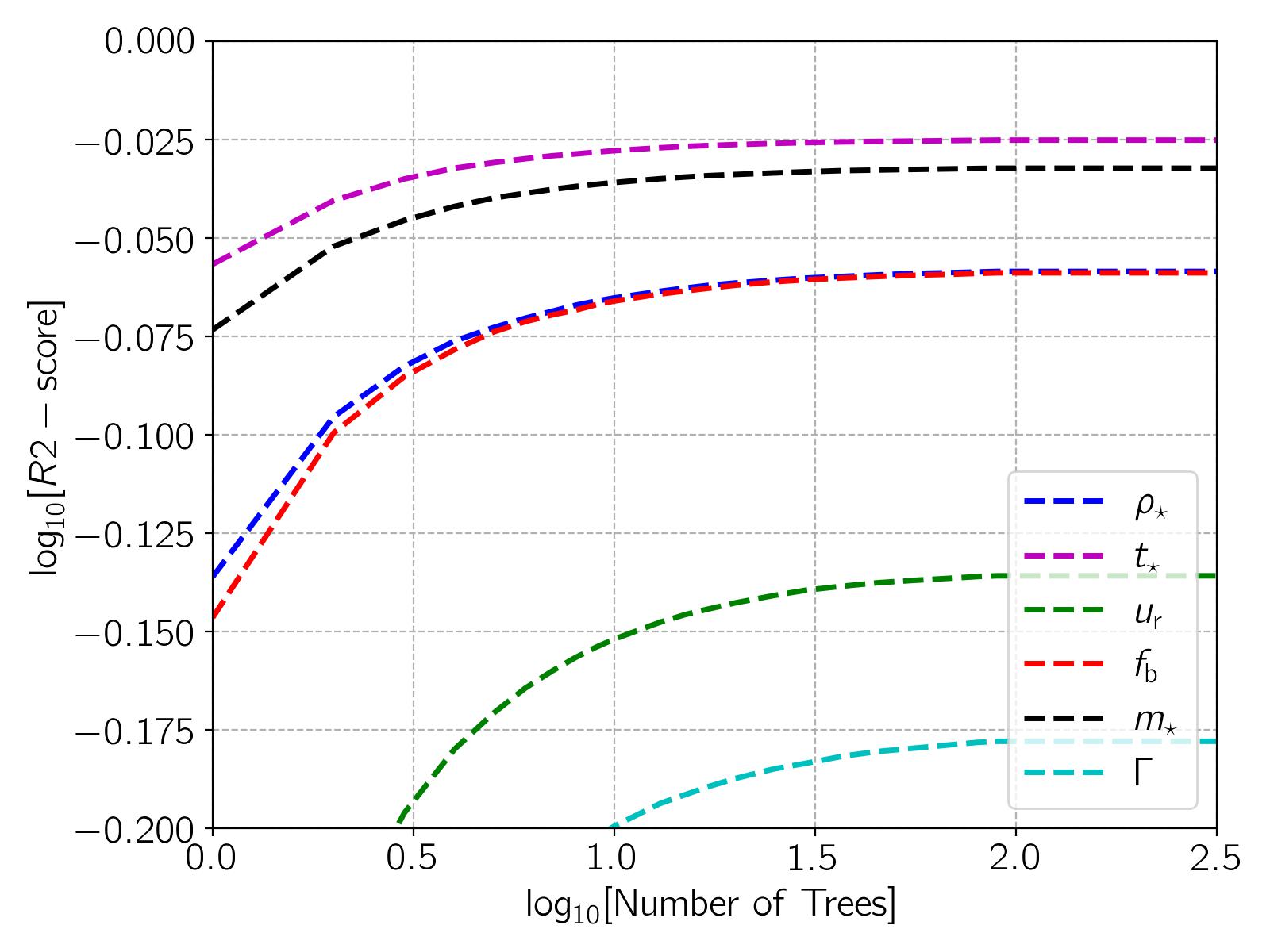}
\caption{$R^{2}$-score versus the number of trees in Random Forests (for predicting different targets as specified in legends) in the logarithmic scale for two ensembles of finite-source microlensing events from uniform (left panel) and limb-darkened (right panel) source stars.}\label{forest}
\end{figure*}

\section{Random Forest approach}\label{tree}
Nowadays, there are several numerical platforms that find the best-fitted models to observational data of microlensing observations by evaluating $\chi^{2}$ maps and searching its local minima \citep[see, e.g.,][]{2007MNRASDominik,2018MNRASBozza,2017AJBachelet,2019ACpoleski}. If there is a continuous degeneracy in microlensing events, the minimum point of the map of $\chi^{2}$ extends to an area. The size of that area determines the errors in the fitted lensing parameters \citep[][]{2010arXivHogg}. This method is the best one can be done for analyzing microlensing observations.
\begin{table*}
\centering
\begin{tabular}{cccccccccc}\toprule[1.5pt]
$\rm{Output}$& $R^{2}-\rm{score}$ & $\rm{MAPE}$&	$\rm{MSE}$& $\rm{RMSE}$& $\rm{IM}_{m_{\rm{base}}}$& $\rm{IM}_{\Delta A}$ & $\rm{IM}_{\rm{FWHM}}$& $\rm{IM}_{\Delta_{\rm{TH}}}$ & $\rm{IM}_{T_{\rm{max}}}$\\
\toprule[1.5pt]
$\rho_{\star}$&$0.860$ &   $0.105$  &   $0.736$  &  $0.858$ & $0.086$ & $\mathbf{0.628}$ &  $0.193$ &  $0.038$ & $0.056$   \\
$t_{\star}$   &  $0.973$ &   $0.051$  &   $5.392$  &  $2.322$ & $0.011$ & $0.136$ &  $\mathbf{0.715}$ &  $0.003$ & $0.136$  \\
$u_{\rm r}$  & $0.769$ &   $2.371$  &   $0.019$  &  $0.139$ & $0.075$ & $0.171$ &  $0.223$ &  $0.028$ & $\mathbf{0.503}$  \\
$f_{\rm b}$  &$0.844$ &   $0.250$  &   $0.013$  &  $0.114$ &$0.145$ & $\mathbf{0.646}$ &  $0.097$ &  $0.049$ & $0.063$  \\
$m_{\star}$  & $0.909$ &   $0.015$  &   $0.122$  &  $0.350$ & $\mathbf{0.744}$ & $0.175$ &  $0.041$ &  $0.013$ & $0.027$  \\
$\rm{Average}$&$0.871$ &  $0.558$  &  $1.256$ &  $0.756$ & $0.212$ & $\mathbf{0.351}$ & $0.254$& $0.026$& $0.157$ \\
\hline
$\rm{Multi}-\rm{output}$ &   $0.843$ &   $0.680$  &   $1.592$  & $1.262$ &$0.022$ & $0.146$ &  $\mathbf{0.701}$ &  $0.005$ & $0.126$  \\
\hline
\end{tabular}
\caption{The $R^{2}$-score, Mean Absolute Percentage Error (MAPE), Mean Squared Error(MSE), Root Mean Squared Error (RMSE) from Random Forest models for predicting different lensing parameters (given in the first column). The feature importances ($\rm{IM}_{i}$s) are mentioned in five last columns. The last row shows the results from a multi-output model which predicts all of 5 lensing parameters, simultaneously.}\label{tab1}
\end{table*}
\begin{table*}
\centering
\begin{tabular}{ccccccccccc}\toprule[1.5pt]
$\rm{Output}$& $R^{2}-\rm{score}$ & $\rm{MAPE}$&	$\rm{MSE}$& $\rm{RMSE}$ &  $\rm{IM}_{m_{\rm{base}}}$& $\rm{IM}_{\Delta A}$ & $\rm{IM}_{\rm{FWHM}}$& $\rm{IM}_{\Delta_{\rm{TH}}}$ & $\rm{IM}_{T_{\rm{max}}}$ & $\rm{IM}_{f_{\rm{pl}}}$\\
\toprule[1.5pt]
$\rho_{\star}$& $0.877$ & $0.093$  &  $0.652$  &  $0.808$ & $0.0619$ & $\mathbf{0.544}$ & $0.226$ & $0.079$ & $0.029$ & $0.061$  \\
$t_{\star}$ &   $0.945$ &    $0.071$  &   $11.697$&  $3.420$ & $0.013$ &$0.085$ &$\mathbf{0.725}$ & $0.010$ & $0.040$ & $ 0.127$\\
$u_{\rm r}$ & $0.740$ &   $2.244$  &  $0.022$  &  $0.148$& $0.059$ & $0.069$ & $0.089$ & $0.062$ & $0.076$ & $\mathbf{0.641}$\\
$f_{\rm b}$ & $0.878$ &   $0.218$  &  $0.011$  &  $0.104$ & $0.132$ & $\mathbf{0.618}$ & $0.075$ & $0.069$ & $0.022$ & $0.083$\\  
$m_{\star}$ & $0.931$ &   $0.013$  &  $0.096$  &  $0.309$ &$\mathbf{0.719}$& $0.176$ & $0.038$ & $0.019$ & $0.012$ & $0.036$  \\
 $\Gamma$ & $0.674$ &  $0.086$ &     $0.003$ &  $0.062$ & $0.069$ & $0.079$ & $0.266$ & $0.048$ & $0.237$ &$\mathbf{0.309}$ \\
$\rm{Average}$ & $0.841$  & $0.405$ &  $2.337$ &  $0.849$ & $0.176$ & $0.261$ & $0.243$ & $0.046$ &$0.061$ & $0.212$  \\
\hline
$\rm{Multi}-\rm{output}$ & $0.787$ &   $0.624$   &  $2.396$   &  $1.548$ &$0.021$ & $0.133$ & $\mathbf{0.684}$ & $0.011$ & $0.038$ &  $0.117$ \\
\hline
\end{tabular}
\caption{Same as Table \ref{tab1}, but for microlensing events from limb-darkened source stars.}\label{tab2}
\end{table*}

Nevertheless and by considering a rapid enhancement of Machine Learning applications in the data science, I study if its approaches can do modeling processes of microlensing data. In this regard, I first make a lot of synthetic finite-source microlensing light curves. Then, I numerically calculate observational parameters for each light curve. I use a suitable ML approach to make a model and find the relations between observational parameters (inputs) and lensing parameters (outputs). Then, that model can predict lensing parameters for an unseen set of observational parameters. However, a more straightforward method is using artificial neural networks, in which we do not need to extract observational parameters from light curves and the machine's inputs are time series data \citep[see, e.g., ][]{2022AJZhao}.

Machine Learning has several types, but two types of them are extensively used, i.e., supervised and unsupervised learning. In different fields of astrophysics, supervised ML is much more applicable \citep[see, e.g., ][]{2021AJKhakpash,2022MNRASexoplanet,2023Fatheddin}. It includes two subtypes: Regression and Classification with continuous and discrete (limited number) outputs, respectively. Some of famous approaches in supervised machine learning are: K-Nearest Neighbor (KNN), Decision Tree, Naive Bayes classifier, Logistic Regression, Support-Vector Machine, Linear Discriminant Analysis, and Artificial Neural Networks. Among them, the most suitable model for our data is not Bayesian-based methods, because our observational parameters (inputs) have strong correlations (see Figure \ref{corri}). Also classification methods are not applicable here, because lensing parameters have continuous amounts. We apply Decision Tree Regression approach from \texttt{scikit-learn} Python module\footnote{\url{https://scikit-learn.org/stable/}}. To improve the model we make Random Forests with many trees and use K-Fold Cross Validation (with 10 Folds) for choosing training sets.
\begin{figure}
\includegraphics[width=0.49\textwidth]{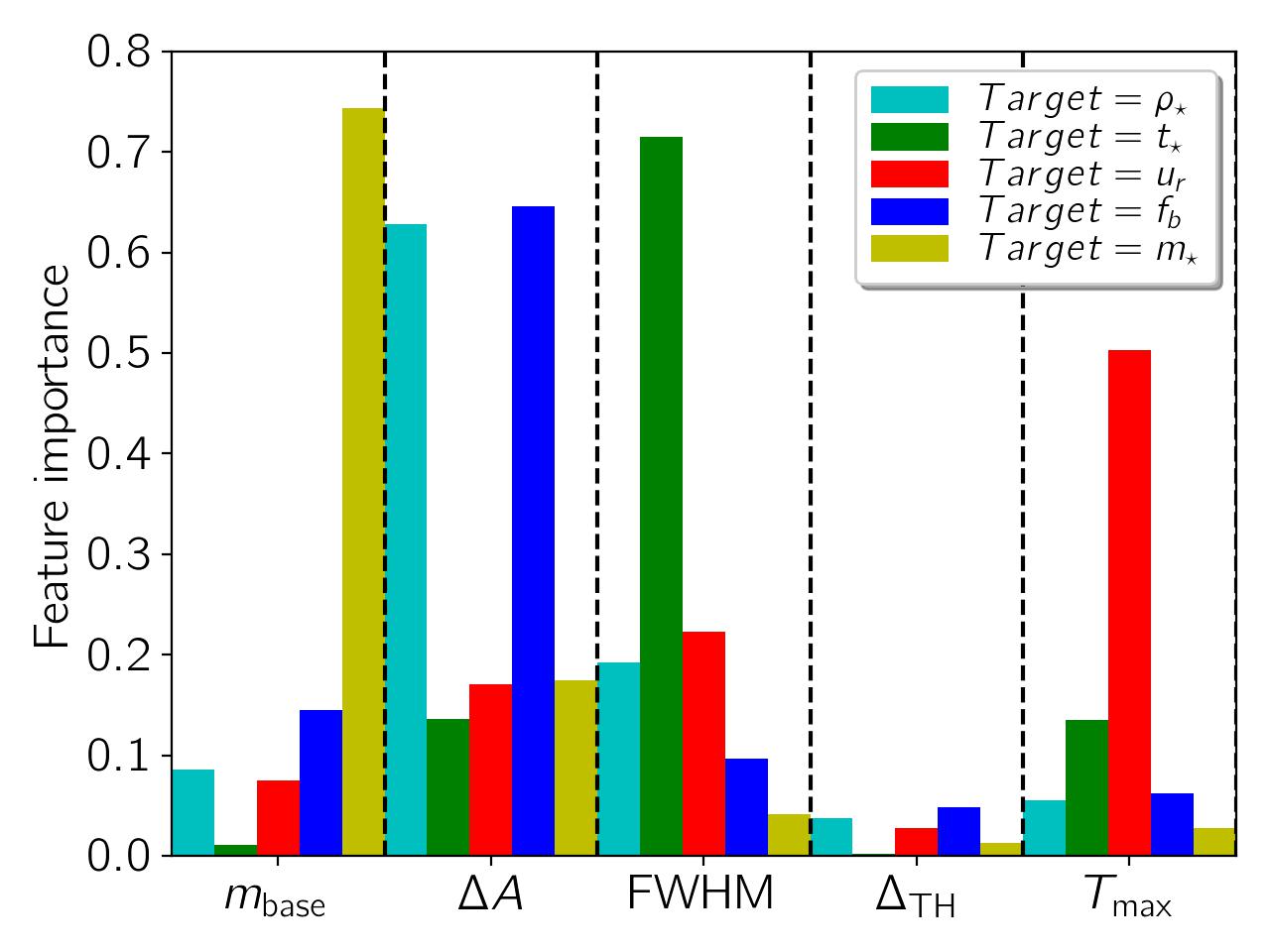}
\includegraphics[width=0.49\textwidth]{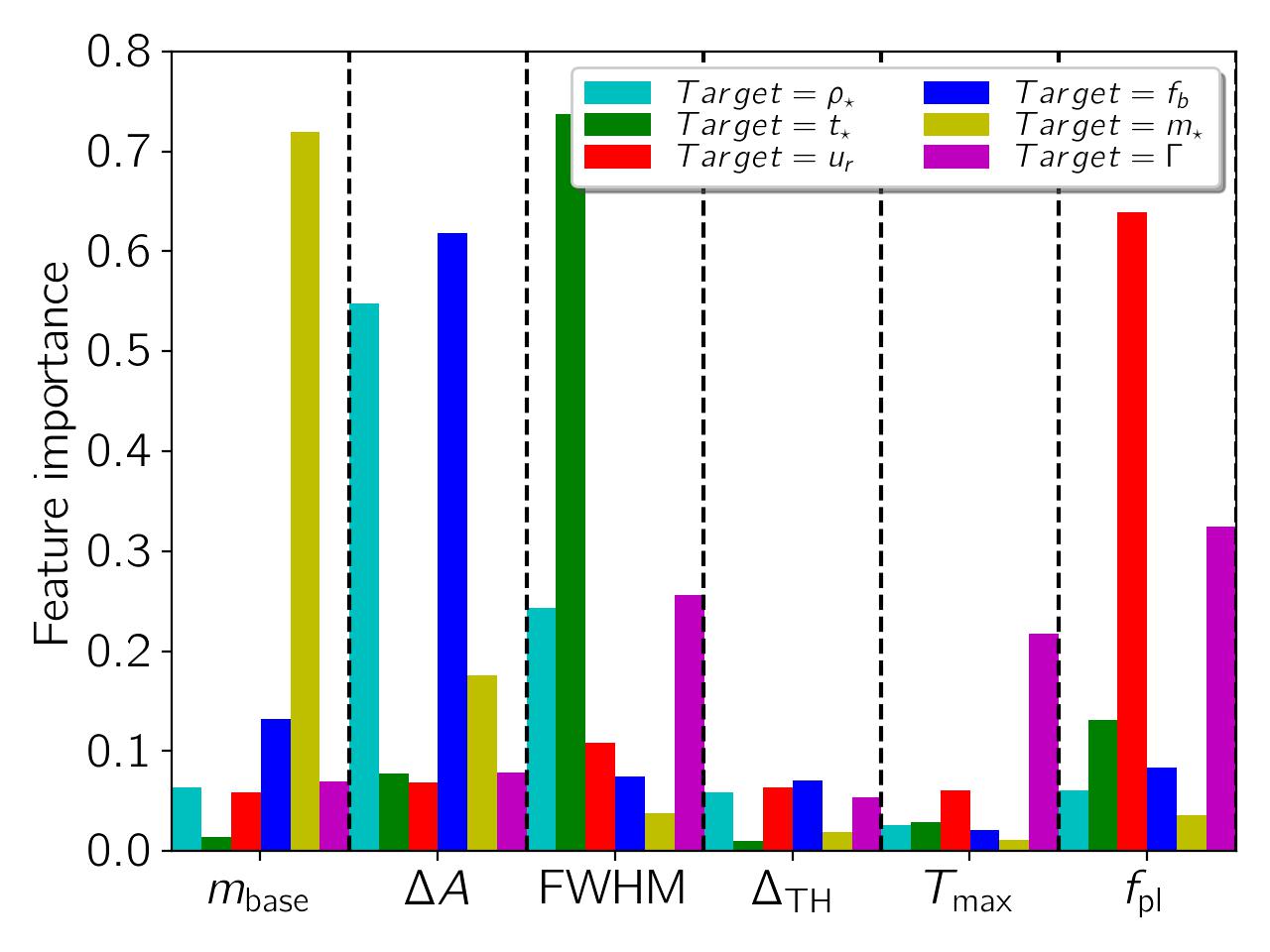}
\caption{The importance of observational parameters (inputs) while predicting lensing parameters in finite-source microlensing events from uniform (top panel) and limb-darkened (bottom panel) source stars.}
\label{import}
\end{figure}

Finding each lensing parameter ($\rho_{\star}$, $t_{\star}$, $u_{\rm r}$, $f_{\rm b}$, $m_{\star}$, and $\Gamma$) separately and using a discrete Random Forest model has a better overall accuracy than finding all lensing parameters from one multi-output Random Forest model. Hence, we make 5 (and 6 for limb-darkened source stars) Random Forest models to predict every lensing parameter separately.  

In order to find the sufficient number of trees for these Random Forests (each one with a specified output), in Figure \ref{forest} we plot the $R^{2}$-score versus the number of trees, for ensembles of finite-source microlensing events from uniform (left panel) and limb-darkened (right panel) source stars. The best numbers of trees for Random Forest models are $\sim 100$. For limb-darkened source stars, the suitable numbers of trees are $\sim 120$, as well. 

In Table \ref{tab1}, we report $R^{2}$-score, Mean Absolute Percentage Error (MAPE), Mean Squared Error (MSE), Root Mean Squared Error (RMSE), from applying models to the test set. Accordingly, $t_{\star}$ will be measured with the highest $R^{2}$-score of $0.97$, and $u_{\rm r}$ will be determined with the least score of $0.77$. The average score for measuring these 5 lensing parameters separately is $0.87$, whereas a multi-output Random Forest model will determine all lensing parameters simultaneously with the $R^{2}$-score of $0.84$. In the five last columns of this table, we report the importance of different inputs for extracting lensing parameters. For each lensing parameter, the highest value of feature importances is highlighted. Hence, $T_{\rm{max}}$ and FWHM have most significant roles to determine $u_{\rm r}$, and $t_{\star}$, respectively. For uniform source stars, $\Delta_{\rm{TH}}$ is very small and is not an important input for extracting any lensing parameter.

I apply K-Fold Cross-Validation Random Forest models to six observational parameters due to finite-source microlensing events of limb-darkened source stars. The results from these single-output and multi-output models are represented in Table \ref{tab2}. Accordingly, the worst $R^{2}$-score is for evaluating the limb-darkening coefficient (see its weak correlations with observational parameters in Figure \ref{corri}). Although $u_{\rm r}$ has simultaneously correlations with three observational parameters FWHM, $T_{\rm{max}}$, and $f_{\rm{pl}}$, in the simulation $f_{\rm{pl}}$ has the highest importance while predicting $u_{\rm r}$. We note that $f_{\rm{pl}}$ is defined by using $T_{\rm{max}}$ (see Equation \ref{fpl}).

Generally, limb-darkening breaks the degeneracy between $T_{\rm{max}}$, and FWHM (compare Figures \ref{uhm1} and \ref{uhm2}), but on the other hand the limb-darkening coefficient and $u_{\rm r}$ have similar effects on microlensing light curves. Also, the limb-darkening itself has weak correlations with observational parameters. For these reasons, the Random Forest model for measuring this lensing parameter has the worst $R^{2}$-score ($0.67$). Hence, limb-darkening has a two-fold effect while modeling finite-source microlensing events.
\begin{figure}
\includegraphics[width=0.49\textwidth]{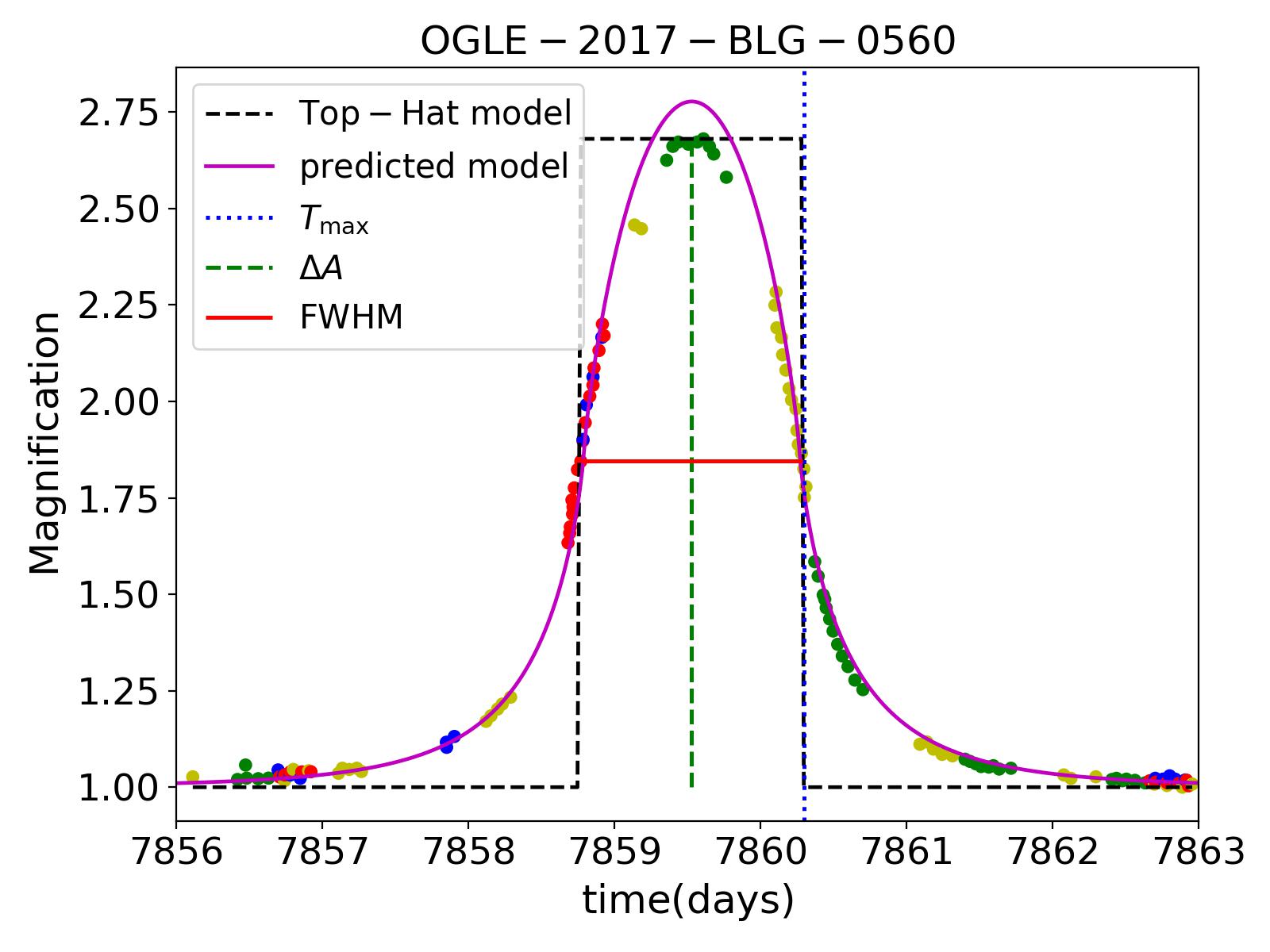}
\caption{The observational data points for the microlensing event OGLE-2017-BLG-0560, which were taken from Fig. (1) of \citet{2019MrozAA}. The observational parameters, including FWHM, $\Delta A$, $T_{\rm{max}}$, are shown with solid red, dashed green, and dotted blue lines, respectively. The top-hat model for this light curve is depicted with dashed black lines.}
	\label{case}
\end{figure}

To better compare importance of different observational parameters while estimating lensing parameters (which are mentioned in last columns of Tables \ref{tab1} and \ref{tab2}), we plot the feature importances to predict different lensing parameters of simulated light curves without (top panel) and with (bottom panel) limb-darkening effects in Figure \ref{import}. Accordingly,  most important inputs are FWHM and $m_{\rm{base}}$ which directly determine $t_{\star}$, and $m_{\star}$, respectively. 

\subsection{Case study:~OGLE-2017-BLG-0560}
For real observational data, extracting observational parameters from  microlensing light curves is not always easy. The accuracy in observational parameters depends on how much data covers light curves. To evaluate this point, for an example short-duration and finite-source microlensing event, i.e., OGLE-2017-BLG-0560 \citep{2019MrozAA}, I extract the observational parameters from data points. In Figure \ref{case}, the observational data points (taken from Figure 1 of \citet{2019MrozAA}), and the measured observational parameters are shown. These parameters are $m_{\rm{base}}=14.29$ mag, $\Delta A=1.68$ (depicted with dashed green line in Figure \ref{case}), FWHM$=1.52$ days (solid red line), $\Delta_{\rm{TH}}=0.08$ (the top-hat model is shown with dashed black lines), and $T_{\rm{max}}=0.77$ days (with respect to $t_{0}$ which is depicted with dotted blue line), and $f_{\rm{pl}}=0.51$. The Random Forest models predict the lensing parameters for this event as $\rho_{\star}=0.79$, $t_{\rm E}=1.18$ days, $u_{0}=0.28$, $f_{\rm b}=1$, $m_{\rm{base}}=14.92$ mag, and $\Gamma=0.47$. I also plot the light curve due to these predicted lensing parameters in Figure \ref{case} with magenta color. Although the predicted model is not fitted to the data very well, but the predicted lensing parameters are close to the best-fitted parameters reported in Table (1) of \citet{2019MrozAA}. Hence, ML models can first predict the initial values of the lensing parameters, and then the best-fitted model will be extracted by searching the $\chi^{2}$ map around those initial values.

\section{Conclusions}\label{conclu}
A continuous degeneracy in extreme finite-source microlensing events exists which prevents us from uniquely specifying the relevant lensing parameters. In this work, I studied the origin of this degeneracy numerically and by evaluating dependency of observational parameters of finite-source microlensing light curves to the lensing parameters. 

A finite-source microlensing light curve from a uniform source star is a function of five lensing parameters, which are $u_{0}$, $t_{\rm E}$, $\rho_{\star}$, $f_{\rm b}$, and $m_{\star}$. In order to measure all of them, we need five independent observational parameters extracted from light curves. These observational parameters are (i) the baseline apparent magnitude $m_{\rm{base}}$, (ii) the maximum enhancement in the magnification factor $\Delta A$, (iii) the Full Width at Half Maximum of light curves (FWHM=$2~t_{\rm{HM}}$), (iv) the time of maximum time-derivative of magnification factor $T_{\rm{max}}$, and (v) the deviation from a top-hat model $\Delta_{\rm{TH}}$. 

Using numerical calculations, I found $T_{\rm{max}} \simeq t_{\rm{HM}}\simeq t_{\star}$, and $\Delta_{\rm{TH}}\to 0$ for $\rho_{\star} \gtrsim 10$ (extreme finite-source events). As a result, for these events a continuous degeneracy appears because the number of independent observational parameters is three. 

By including the limb-darkening effect, $t_{\rm{HM}}$ depends additionally on the limb-darkening coefficient, and it does not tend to $t_{\star}$ even for large source radii. On the other hand, the limb-darkening effect and the lens impact parameter have similar impacts on light curves. Hence, limb-darkening has a two-fold effect in finite-source microlensing light curves. 

To evaluate how much these observational parameters are efficient to specify the lensing parameters, I used numerical approaches, i.e., Random Forests from Decision Trees, and applied these models to a large sample of finite-source microlensing events from uniform source stars. These models predicted lensing parameters from the observational parameters of simulated light curves with an  average $R^{2}$-score of $0.87$. $t_{\star}$, and $u_{\rm r}$ were determined with the highest and lowest scores $0.97$, and $0.77$, respectively. Also, I found that among all lensing parameters FWHM and $m_{\rm{base}}$ have had the highest importances ($\simeq 0.7$) while modeling, because they have directly extracted $t_{\star}$, and $m_{\star}$, respectively.     

I considered the limb-darkening effect for source stars, by adding another lensing parameter $\Gamma$. Hence, the number of observational parameters enhanced by one, and $f_{\rm{pl}}$ was added (given in Equation \ref{fpl}). I used a similar numerical approach, i.e., six separate Random Forests each contains 120 Decision Trees (and by using K-Fold Cross Validation while making training dats), to predict the lensing parameters from observational parameters. Adding limb-darkening causes (i) $t_{\rm{HM}}$ depends on the limb-darkening coefficient and as a result it is separated from $T_{\rm{max}}$, and (ii) on the other hand $\Gamma$ and $u_{\rm r}$ have similar effects on microlensing light curves. Therefore, it may break degeneracy even for extreme finite-source events, and increases $R^{2}$-scores for predicting $\rho_{\star}$, $f_{\rm b}$, and $m_{\star}$. But, $R^{2}$-scores for predicting $u_{\rm r}$ and $\Gamma$ are low ($\sim 0.7$). The correlations of $\Gamma$ with observational parameters are weak.  

\section*{Acknowledgements}
I thank the anonymous referee for his/her careful and useful comments, which improved the quality of the paper.

\section*{DATA AVAILABILITY}
All simulations that were done for this paper are available at: \url{https://github.com/SSajadian54/Finite_source_Microlensing}

\bibliographystyle{mnras}
\bibliography{paperref}
\appendix
\section{Dependence of observational parameters on lensing parameters}\label{append1}
According to Figure \ref{uhm}, three observational parameters $u_{\rm{HM}}$ (the lens-source distance projected on the lens plane normalized to the Einstein radius at the half maximum), and $\Delta_{\rm{TH}}$ (the average squared relative deviation in the magnification factor from a top-hat model), and $f_{\rm{pl}}$ depend on the lens impact parameter in addition to the normalized source radius. For different values of the lens impact parameter we fit proper functions to them versus $\rho_{\star}$ as given by:
\begin{eqnarray}
u_{\rm{HM}}&=&\rho_{\star} \exp\big(a_1 + a_2 \rho_{\star} + a_3 \rho_{\star}^{2}\big )+ a_4 \rho_{\star}, \nonumber \\
\Delta_{\rm{TH}}&=&b_1 \tan^{-1}\big(b_2 \log_{10}[\rho_{\star}] +b_3\big)+b_4),\nonumber \\
f_{\rm{pl}}&=&c_{1} \sqrt{\rho_{\star}} + c_{2} \rho_{\star} +c_{3} \rho_{\star}^{2} +c_{4},
\label{uhmd}
\end{eqnarray}
where, the coefficients $a_{i}$, $b_{i}$, and $c_{i}$ are functions of the lens impact parameter. We calculate these coefficients and report them in Table \ref{app1}, Table \ref{app2}, and Table \ref{app3}, respectively.
\begin{table}
\centering
\begin{tabular}{ccccc}\hline 
$u_{\rm r}$& $a_{1}$ & $a_{2}$&$a_{3}$& $a_{4}$ \\
\toprule[1.5pt]
$0.00$ & $-1.915$ & $-2.588$ & $-0.285$ & $0.996$\\
$0.05$ & $-1.912$ & $-2.585$ & $-0.277$ & $0.996$\\
$0.10$ & $-1.902$ & $-2.577$ & $-0.251$ & $0.997$\\
$0.15$ & $-1.885$ & $-2.565$ & $-0.209$ & $0.997$\\
$0.20$ & $-1.862$ & $-2.547$ & $-0.154$ & $0.997$\\
$0.25$ & $-1.832$ & $-2.524$ & $-0.091$ & $0.997$\\
$0.30$ & $-1.796$ & $-2.493$ & $-0.023$ & $0.997$\\
$0.35$ & $-1.754$ & $-2.454$ & $0.044$ & $0.997$\\
$0.40$ & $-1.706$ & $-2.406$ & $0.105$ & $0.998$\\
$0.45$ & $-1.648$ & $-2.367$ & $0.193$ & $0.998$\\
$0.50$ & $-1.595$ & $-2.248$ & $0.176$ & $0.998$\\
$0.55$ & $-1.536$ & $-2.125$ & $0.160$ & $0.998$\\
$0.60$ & $-1.472$ & $-2.000$ & $0.146$ & $0.999$\\
$0.65$ & $-1.401$ & $-1.876$ & $0.133$ & $1.000$\\
$0.70$ & $-1.323$ & $-1.751$ & $0.121$ & $1.001$\\
$0.75$ & $-1.239$ & $-1.628$ & $0.110$ & $1.002$\\
$0.80$ & $-1.147$ & $-1.501$ & $0.099$ & $1.003$\\
$0.85$ & $-1.046$ & $-1.373$ & $0.087$ & $1.005$\\
$0.90$ & $-0.933$ & $-1.243$ & $0.076$ & $1.008$\\
$0.95$ & $-0.806$ & $-1.111$ & $0.064$ & $1.015$\\
\hline
\end{tabular}
\caption{The coefficients of the best exponential models (given in Eq. \ref{uhmd}) fitted to $u_{\rm{HM}}$ versus $\rho_{\star}$ for different values of $u_{\rm r}$, as plotted in Figure \ref{uhm1}.}\label{app1} 
\end{table}
\begin{table}
\centering
\begin{tabular}{ccccc}\hline
$u_{\rm r}$& $b_{1}$ & $b_{2}$&$b_{3}$& $b_{4}$ \\
\toprule[1.5pt]
$0.00$ & $-0.328$ & $5.964$ & $3.390$ & $0.469$\\
$0.05$ & $-0.332$ & $5.969$ & $3.391$ & $0.475$\\
$0.10$ & $-0.337$ & $5.971$ & $3.392$ & $0.482$\\
$0.15$ & $-0.342$ & $5.973$ & $3.394$ & $0.489$\\
$0.20$ & $-0.347$ & $5.977$ & $3.396$ & $0.497$\\
$0.25$ & $-0.352$ & $5.981$ & $3.399$ & $0.504$\\
$0.30$ & $-0.357$ & $5.986$ & $3.402$ & $0.512$\\
$0.35$ & $-0.363$ & $5.992$ & $3.406$ & $0.519$\\
$0.40$ & $-0.368$ & $5.998$ & $3.410$ & $0.527$\\
$0.45$ & $-0.374$ & $6.005$ & $3.415$ & $0.536$\\
$0.50$ & $-0.380$ & $6.013$ & $3.419$ & $0.544$\\
$0.55$ & $-0.386$ & $6.020$ & $3.424$ & $0.553$\\
$0.60$ & $-0.392$ & $6.028$ & $3.429$ & $0.562$\\
$0.65$ & $-0.399$ & $6.036$ & $3.435$ & $0.571$\\
$0.70$ & $-0.406$ & $6.044$ & $3.440$ & $0.581$\\
$0.75$ & $-0.413$ & $6.051$ & $3.444$ & $0.591$\\
$0.80$ & $-0.420$ & $6.058$ & $3.448$ & $0.602$\\
$0.85$ & $-0.428$ & $6.064$ & $3.452$ & $0.613$\\
$0.90$ & $-0.436$ & $6.068$ & $3.455$ & $0.624$\\
$0.95$ & $-0.444$ & $6.071$ & $3.456$ & $0.636$\\
\hline
\end{tabular}
\caption{The coefficients of the arctangent models (given in Eq. \ref{uhmd}) which are fitted to $\Delta_{\rm{TH}}$ versus $\log_{10}\big[\rho_{\star}\big]$ by considering different values of $u_{\rm r}$ as plotted in Figure \ref{devia}.}\label{app2}
\end{table}

\begin{table}
\centering
\begin{tabular}{ccccc}\hline
$\Gamma$& $c_{1}$ & $c_{2}$&$c_{3}$& $c_{4}$ \\
\toprule[1.5pt]
$0.40$ & $0.267$ & $-0.068$ & $0.002$  & $0.356$\\
$0.42$ & $0.268$ & $-0.067$ & $0.002$  & $0.358$\\
$0.44$ & $0.269$ & $-0.067$ & $0.002$  & $0.360$\\
$0.46$ & $0.270$ & $-0.067$ & $0.002$  & $0.362$\\
$0.48$ & $0.271$ & $-0.066$ & $0.002$  & $0.364$\\
$0.50$ & $0.270$ & $-0.065$ & $0.001$  & $0.366$\\
$0.52$ & $0.271$ & $-0.065$ & $0.001$  & $0.368$\\
$0.54$ & $0.271$ & $-0.065$ & $0.001$  & $0.371$\\
$0.56$ & $0.272$ & $-0.064$ & $0.001$  & $0.373$\\
$0.58$ & $0.273$ & $-0.064$ & $0.001$  & $0.376$\\
$0.60$ & $0.274$ & $-0.063$ & $0.001$  & $0.378$\\
$0.62$ & $0.274$ & $-0.063$ & $0.001$  & $0.381$\\
$0.64$ & $0.275$ & $-0.063$ & $0.001$  & $0.383$\\
$0.66$ & $0.276$ & $-0.062$ & $0.001$  & $0.386$\\
$0.68$ & $0.276$ & $-0.062$ & $0.001$  & $0.388$\\
$0.70$ & $0.277$ & $-0.061$ & $0.001$  & $0.391$\\
$0.72$ & $0.278$ & $-0.061$ & $0.001$  & $0.393$\\
$0.74$ & $0.279$ & $-0.061$ & $0.001$  & $0.396$\\
$0.76$ & $0.279$ & $-0.060$ & $0.001$  & $0.399$\\
$0.78$ & $0.281$ & $-0.060$ & $0.001$  & $0.401$\\
\hline
\end{tabular}
\caption{The coefficients of best power-law models fitted to $f_{\rm{pl}}$ (given in Eq. \ref{fpl}) versus $\rho_{\star}$ for different values of $\Gamma$, and $u_{0}=0$ as depicted in Figure \ref{fplfig}.}\label{app3}
\end{table}
\end{document}